\documentclass{article}
\usepackage{epsfig}
\def\aprle{\buildrel < \over {_{\sim}}} 
\def\aprge{\buildrel > \over {_{\sim}}} 

\begin{document}     

\title{The geometry of   atmospheric neutrino production }

\author{Paolo Lipari  \\
Dipartimento di Fisica, Universit\`a di Roma ``la Sapienza",\\
and I.N.F.N., Sezione di Roma, P. A. Moro 2,\\ I-00185 Roma, Italy \\
also at: Research Center for Cosmic  Neutrinos, \\
 ICRR, University of Tokyo \\
Midori--cho 3--2--1, Tanashi-shi, Tokyo 188-8502, Japan}

\date{February 27, 2000}          
\maketitle

\large

\begin{abstract}
The zenith angle distributions of atmospheric neutrinos  are determined
by the  possible presence of  neutrino oscillations
and the combination of three most important contributions: (1) geomagnetic
effects on the primary cosmic rays,
that  suppress
the primary flux  in the Earth's magnetic  equatorial region,
 (2) the zenith angle dependence of the
neutrino yields,
due to the fact that inclined   showers
 produce more neutrinos, 
and (3)  geometrical  effects  due to the spherical
shell geometry of the neutrino  production   volume. 
 The last effect has been  recognized
only recently and  results  in  an important enhancement  of the
flux  of  sub--GeV neutrinos  for horizontal directions.
In this  work we   discuss the
geometrical  effect   and its relevance  in the interpretation
of the atmospheric neutrino data.
\end{abstract}

\section{Introduction}
Atmospheric  neutrinos are produced  in the hadronic   showers
generated  by  cosmic   rays in the Earth's  atmosphere.
Absorption in the Earth is  negligible
in the  entire  relevant energy range,
and therefore  a   detector  located near  the  surface of the Earth
will receive  a  neutrino flux from all directions.
In   the presence   of neutrino oscillations the  observed rates of
$\nu$ interactions   will be  modified, and the
angular  distributions  distorted.
These effects  have been in  fact  measured by Super--Kamiokande
\cite{SK} and other atmospheric  neutrino detectors 
\cite{Kamiokande,IMB,Soudan,MACRO}
and give  clear  evidence for the existence  of
flavor  transitions
.%$\nu_\mu \leftrightarrow \nu_\tau$ oscillations
%with  large  mixing  and  
%$|\Delta m^2|  \simeq 2$--$6 \times 10^{-3}$~eV$^2$.

The evidence for oscillations is robust,  and does not depend
on a detailed  calculation of the  expected  neutrino  fluxes
with and without neutrino oscillations.
The   strongest  evidence  comes  in fact  from 
the detection of an   up--down  asymmetry
in the angular  distribution  of  muon  events 
and in  a   small ratio  for the  rates  of
$\mu$--like  and $e$--like events.
It is  possible to  predict with very simple  considerations
that in the absence of oscillations the $\nu$  fluxes are approximately
up--down symmetric, and the fluxes of  electron and muon neutrinos
are  strictly related  because they are   produced in the chain decay
of the same parent  mesons. 
In order to extract the oscillation parameters from the data
it is however important  to have   detailed  predictions for the
expected  intensity and angular distributions of the
no--oscillation fluxes.

In this work  I will  discuss the 
angular  distributions of  the atmospheric  neutrino  fluxes.
These distributions are  determined by the
presence of neutrino oscillations (if they exist) and
a combination of three  most important effects:

\begin {enumerate}

\item Geomagnetic effects on the primary cosmic rays.
Low  rigidity  particles  cannot  reach  the vicinity of
the Earth, and the  effect  depends  on the  point of the Earth 
where  cosmic  rays  arrive  (being stronger  near the geomagnetic
equator), and on their  direction  (local zenith  and azimuth angles).

\item The  zenith angle  dependence of the neutrino yields.
Inclined  showers  produce  more  neutrinos  that  vertical ones
because the decay of charged  mesons  and muons is  then  more
probable.

\item The spherical geometry of the neutrino source volume.
This  results  in an enhancement of the  neutrino fluxes
from  horizontal directions  and a suppression for the
(up--going and down--going) vertical directions.
The effect is exactly up--down  symmetric, and is  important 
for  sub--GeV  neutrinos.

\end{enumerate}

Additional  smaller  effects are 
due to  the bending of charged  particles 
during  shower development, the presence of mountains,
and the  existence of  different
air density profiles  over  different  geographycal  regions.

The aim  of this paper is to  discuss in particular
 the  effect of the spherical geometry
of the  neutrino  source region.
This  effect  has  been  recognized  only recently
\cite{fluka-3d},  is not included in the calculations
currenty in  use \cite{HKKM,bartol} in the analysis  of experimental data
and is still the object of some controversy.

The  work is organized as follows:
the next section  contains a   discussion
what is the one--dimensional  (1D)  approximation 
used in the  first  calculations of the atmospheric  neutrino fluxes
and  why  the approximation was  considered   necessary;
section 3 and  4   present   for  completeness
 very brief  qualitative discussion  on  the 
geomagnetic  effects  and 
of the  angular dependence of the neutrino  yields;
section 5 discusses the  geometrical  effects 
related to the shape of the 
neutrino source  volume;
section 6  presents the results of a   detailed  calculations;
section 7 gives  a summary.

\section {The one--dimensional  approximation}
Any cosmic  ray  shower,
produced  by a primary  particle 
interacting  in  an arbitrary point  of the  atmosphere,  with
an  arbitrary direction can
result in   one (or more) neutrinos   with  trajectories 
that intersect the
detector  in consideration. 
Of course only  a very small  fraction of the  cosmic  ray 
showers   produce  neutrinos   that pass inside (or in the vicinity)
of a detector, and therefore a   full
montecarlo calculation of the neutrino  fluxes 
appears as extremely inefficient.

It is  because of this  difficulty   that   the first generation 
calculations  of the atmospheric neutrino  fluxes  have been performed
in what is called the one--dimensional approximation.
In this  approximation  one considers  only a  very small  subset of 
all possible  cosmic rays   primaries, those  that have a  trajectory
that  when   continued  as  a straight line  beyond the interaction point
intersects  the detector. The neutrinos  produced  in the showers
of these primaries are  considered as  collinear to the  primary parent, 
and  also  have  trajectories  that  intersect the detector,
therefore  all  $\nu$'s   generated in  a montecarlo calculation can be
collected and  analysed  to  estimate  the atmospheric  neutrino fluxes.

More explicitely, in a  1--D  calculation the 
flux  of neutrinos  with  flavor $\alpha$, 
energy $E_\nu$  and  direction  $\Omega$
observed  by a  detector  located   at the position
$\vec{x}_{d}$  is   calculated as:
\begin{equation}
\phi_{\nu_\alpha} (E_\nu, \Omega_\nu, \vec{x}_d) =
\sum_A ~\int dE_0 ~ \phi_A[E_0, \Omega_0(\Omega_\nu), \vec{x}_0(\Omega_\nu,
\vec{x}_d)]~ {dn_{A\to \nu_\alpha} \over dE_\nu} (E_\nu; ~E_0, \cos \theta_0)
\label{eq:1d}
\end{equation}
the sum  is over all primary  cosmic rays nuclear species,
the quantity $dn_{A \to \nu_\alpha}/dE_\nu$  is the  differential  yield
of  neutrinos  of flavor $\alpha$    from a primary of 
type $A$, energy $E_0$ and zenith angle $\theta_0$, 
$\phi_A(E_0, \Omega_0, \vec{x}_0)$ 
is  the flux  of  primary  cosmic  rays of 
type  $A$ ,  energy $E_0$ and direction  $\Omega_0$   that 
reach the  Earth at  point  $\vec{x}_0$.
The  dependences on  the direction and  position
for  the primary flux
are the consequence of  geomagnetic effects.
Since  neutrinos  travel along straight lines, 
the interaction position of the  primary  flux  is  determined
to a  good  approximation by the    neutrino
direction $\Omega_\nu$. 
All   down--going  neutrinos  ($\cos \theta_\nu > 0$)  are
produced   in   showers  in the  general vicinity of the detector:
$\vec{x}_0 \simeq  \vec{x}_d$, while  for  up--going 
trajectories ($\cos \theta_\nu < 0$) the primary interaction point is
in the  vicinity of the point where the neutrino enters the Earth.
The   crucial  approximation made in the 1--D approximation, is  to
consider also  the direction of the primary  $\Omega_0$   as
  determined  by $\Omega_\nu$.
For  down--going  neutrinos  this   simply means 
$\Omega_0 \simeq \Omega_\nu$,   and for  up--going
neutrinos,
expressing $\Omega_0$ as  a zenith and  azimuth angle  with respect to the
local vertical,  this  means:
$\Omega_0 \equiv (\cos\theta_0, \varphi_0) 
= (-\cos\theta_\nu, \varphi_\nu)$.

\begin{figure} [t]
\centerline{\psfig{figure=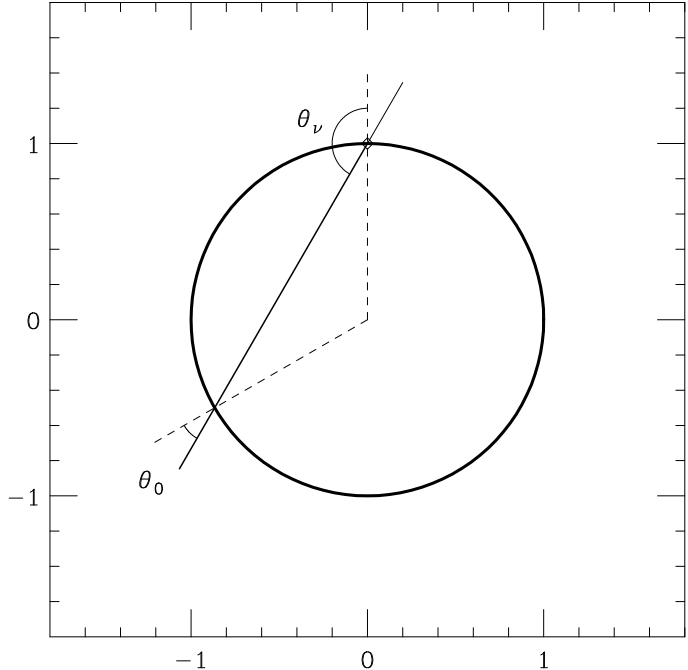,angle=90,height=9.0cm}}
\caption {Relation between  the  zenith  angles 
of the neutrino ($\theta_\nu$)  and the primary  particle
($\theta_0$) in the 1--D approximation.
\label{fig:1d}  }
\end{figure}

Equation (\ref{eq:1d})  simplifies  enormously the 
calculation of the atmospheric  neutrinos  fluxes   because now 
only  a  very small   (indeed  infinitesimal)  fraction
 of  the    cosmic  ray showers 
has  to be studied.
About  half of these  showers have trajectories 
that  reach the Earth  near  the detector  point 
(arriving from  all  directions in  a $2\pi$ solid angle), 
the other  half  is distributed   over the 
entire surface of the Earth, but a  unique  direction   has  to be 
considered  for  each point (see fig.~\ref{fig:1d}).

\vspace {0.3 cm}
In the equation (\ref{eq:1d})    the problem of
the calculation of the neutrino  is ``factorized'' in three parts:
\begin{enumerate}
\item  Determination of the primary  cosmic  ray  fluxes. 
These fluxes  have  of course to be  measured  
experimentally.  
%At low  energy the measurement  has to be
%performed near the Earth magnetic polar regions  where 
%the  geomagnetic effects  do not  forbid the arrival of low  rigidity
%particles.
%This  measurement correspond to the isotropic flux
%of  cosmic rays at one astronomical  unit of   distance from the sun
%uperturbed  by  geomagnetic effects.
The  primary   fluxes  have  also a   time  dependence  due to  the solar
modulation. The effect decreases  with  momentum,  and  becomes
negligible  for rigidities $p/q \aprge  100$~GV.

\item  Calculation  of  the geomagnetic  effects.
The  primary cosmic  ray fluxes  reaching point 
$\vec{x}_0$  from the direction $\Omega_0$ can be  calculated 
from  the  isotropic flux  unperturbed
by geomagnetic  effects  
(that is measured in the Earth's magnetic polar regions)
with a  knowledge  of the structure of the  geomagnetic  field, testing
if the trajectories are allowed  or  forbidden.

\item  Calculation of the neutrino yields.
The number of  neutrinos  (and  their  energy spectrum)
produced  by a primary of energy $E_0$  and   zenith angle $\theta_0$
can  be calculated   studying the average development 
of  hadronic  showers in the atmosphere. 
\end{enumerate}

A calculation using  a three--dimensional  method
can  still be considered  as    divided into  the three parts discussed
above,  and indeed the first two  steps  are identical. 
The  calculation  is   however much more complex
because  it is not  anymore  sufficient 
to  consider for the  neutrino  yield simply the inclusive
energy distribution of the   neutrinos  produced in a shower, but also
the angular  distributions  of  the neutrinos (and  their  strong correlation
with the energy) are  important and have to be calculated.
Moreover the simple integral 
in  (\ref{eq:1d})  performed   over
the primary energy $E_0$  keeping  the   direction
fixed  has now to be replaced    by a more complex
convolution   that involves   not only the
energy but also the direction of the  primary
particle, since  the   neutrinos 
with  a   given trajectory can be produced
in showers with a non collinear axis.

\section {Geomagnetic  effects}
An assumption common to all atmospheric  neutrino calculation,   is that
the fluxes of  cosmic  rays  at one astronomical unit of   distance from
the sun, when  they are  not perturbed  by the  near 
presence of the Earth are isotropic, and can be  simply
  described by their  energy
dependence $\phi_A(E_0)$.
The fluxes reaching the  surface of the Earth are however
affected by the geomagnetic field and therefore are not isotropic,
and have different  intensities   in different locations. For example
it is  intuitively clear  that  low  rigidity particles
can  reach the Earth  only  traveling  parallel  to the
magnetic field  lines  arriving  near the  magnetic poles,
while  high rigidity particles can reach   all points on the 
Earth's   surface  from all possible directions.
 
Given a map of the magnetic  field 
around the Earth it is a straightforward exercise to  
compute numerically if a given   primary  particle  three--momentum
and position   correspond to an allowed or a forbidden trajectory.
 It is sufficient  to  integrate  the equation of  motions
of the charged  particle  in the  geomagnetic  field, and see if the
past  trajectory of the particle 
intersects  the Earth's surface,  remains  confined
to a finite  distance   from the Earth   (forbidden  trajectory);
or  originates from large   distances  (allowed trajectory).

For a qualitative  understanding it can  be useful  to consider the
historically  important  approximation of    describing the 
Earth geomagnetic  effect as  an exactly dipolar field.
In a dipolar field   the  problem can  be solved  analytically.
All   positively    particles with   rigidity
$R > R_S^+$  are allowed
and the  trajectories of all particles  with  
$R < R_S^+$  are forbidden
because they remain  confined  to a finite distance from
the dipole center \footnote{In this  solution it is assumed
that  the field  fills  the entire space.  A  fraction of the 
allowed  trajectories  have really a   segment 
``inside'' the Earth (with $r < R_\oplus$)  and therefore  should be
considered  forbidden also in a dipole field.}.
The quantity $R_S^\pm(\vec{x}, \hat{n})$
is the St\"ormer    rigidity  cutoff \cite{Stormer}:
\begin {equation}
R_S^+ (r, \lambda_M, \theta, \varphi) =  \left ( {M \over 2 r^2 }
\right )  ~ \left \{  { \cos^4 \lambda_M \over
[1 + (1 - \cos^3 \lambda_M \sin \theta \sin \varphi)^{1/2}]^2 } \right \}
\label{eq:Stormer}
\end{equation}
where we have made use of the cylindrical symmetry  of the problem,
 $M$  is the magnetic dipole moment of  field,
$r$ is the distance from the dipole center, $\lambda_M$ the magnetic
latitude, $\theta$  the zenith angle and
$\varphi$ an  azimuth angle  measured counterclockwise from magnetic 
north. 
For  negatively  charged particles the cutoff
is obtained   with the   reflection 
$\varphi \to \varphi + \pi$,  that  is exchanging 
east  and west.

\vspace {0.3 cm}
The  qualitative  features  that  are important for  our
discussion are the following:

\begin{enumerate}
\item  For  a   fixed direction
the cutoff   rigidity   grows monotonically   from  a
 vanishing  value  at the magnetic pole  to  a  maximum   value
at the magnetic  equator.

\item  The cutoffs for  particles traveling  toward magnetic  west
are  higher than  those for particles  traveling toward magnetic  east.
Note that geomagnetic  effects  are the only   mechanism
discussed in this work  that  can generate a 
non--flatness  in  the azimuthal angle distributions  of 
atmospheric neutrinos.

\item The  highest  cutoff corresponds to 
westward going,  horizontal particles  reaching the 
 surface of the Earth at
the magnetic equator
($\varphi = 90^\circ$, $\theta = 90^\circ$, $\lambda_M = 0^\circ$).
The  maximum    rigidity cutoff is approximately 60~GV. 
All geomagnetic  effects  vanish    for  particles 
above the   rigidity value.
\end{enumerate}

\section {The neutrino  yields}
\label{sec:yield}
The ``neutrino yield''
is  the average number of neutrinos  of  a 
certain flavor produced by a primary cosmic
ray particle. It  depends
on  the primary particle type,  energy, and  zenith angle.
There are also weaker dependences  on the azimuth angle of
the shower and  its  geographycal  locations
due to the  effects of  the geomagnetic  field on the shower
development.

In the  1--D  approximation    the   angles  of the  secondary particles
with respect to the primary directions are  neglected 
(or integrated over),  and therefore   the   neutrino  yields  can
be described  simply  as    a  set of  functions 
that   give  the  energy distribution  of the 
neutrinos  (of  different flavors)  produced  by    primaries of  given
type, energy,  and zenith angle.

The one--dimensional neutrino  yields   have  been  obtained
with   the  integration  (numerical \cite{Volkova,BN,butkevich}
  or even analytical
in  simplified  tratments \cite{lipari}) of a set of differential  equations
that   describe the average development  of hadronic  showers.
The most  accurate  calculations of the yields  are  however  performed
with montecarlo  techniques.  A  large number of  showers   (for
primary particles of different  type,  energy   and  zenith angle)
is generated,  and  the number,   flavor   and energy of all produced
neutrinos is  recorded to compute  numerically  the neutrino yields
as a function of primary energy,  direction and particle  type.

The   authors of the  montecarlo calculations of  the yields
\cite{HKKM,bartol} have also usually made 
the  approximation  of considering
 all secondary particles in the shower  as  collinear to the 
primary particle. This  can   be obtained   ``rotating'' all
final state  particles  so that their momentum 
is  parallel  to the projectile (or parent)  particle, and
neglecting  multiple  scattering  and  bending  in the  geomagnetic field.
Strictly  speaking   this rotation of course implies
a small (negligible)
deviation from exact conservation of the longitudinal momentum.
In the procedure  energy is  exactly conserved.

The neutrino  yields  have some  important dependence on the zenith angle
of the primary  particle.
The  yields    grow  monotonically
when  the  zenith  angle of the 
primary  particle   changes from  the vertical to the horizontal  direction.
The reasons  for this  growth  are simple to  understand,
and   originate  from two  effects.
The  first effect is  simply that an  inclined  shower   develops in air
for a longer distance  before hitting the ground,  therefore the muons 
produced in  high zenith angle  showers  have more time to decay in air
and  produce  more neutrinos.
Essentially all  muons  that hit  the ground rapidly lose  all their
energy in ionization and radiation processes and decay  at rest
producing very soft neutrinos
(with $E_\nu < m_\mu/2$)  that   can be  neglected 
because are   below the  thresholds of   the existing 
detectors.

A second effect, contributing also  to an enhancement 
of neutrino  production  for  horizontal  showers 
is due   to the fact that  the air density is 
not    constant   but      decreases (approximately  exponentially in the 
stratosphere)  with  increasing  altitude.
Therefore the inclination
of a  trajectory  (assuming the same starting point)
determines the column density $X$  (g~cm$^{-2}$) that  corresponds to
a   fixed  length $L$ (cm).
The column density   is  largest  for   vertically
down--going particles (zenith angle $\theta = 0$)  and 
decreases  monotonically   with  increasing  zenith angle.
The relation between  $L$ and $X$  determines the
relative probability
of  decay or  interactions  for  charged pions (or kaons),  
and the  energy loss
of a muon  before  decay.
A smaller  $X$   (for  a fixed $L$) implies  that  the interaction
of weakly decaying   mesons  is suppressed  and their decay  enhanced;
it also  means that the muons  will 
decay with  a higher   energy.  
For  inclined  showers
the first effect    results
in  a higher  number of neutrinos,
and the second one  in a slightly harder spectrum    of neutrinos

The ratio between the vertical  and  horizontal yields  is  a 
function of the neutrino energy.  
For low $E_\nu$ the   ratio is  close to unity, and there is  very little
enhancement  for the horizontal  directions. This  can  be qualitatively
easily understood.  Low  energy neutrinos
are  produced in  the decay   of low  energy pions
and low  energy muons.  
Because of   relativistic  effects the decay length  of  unstable  particles
is  proportional  to    the particle momentum, for low  energy 
particles  the  decay  is  so  
 rapid, that  the decay  probability  
is  unity, and the energy loss before decay is  
negligible   independently from the direction
of the particle.  In  these  circumstances there is no
enhancement for the   horizontal directions, and the yield is  isotropic.
With increasing  energy  the  decay  length  increases,  and the effects
outlined  above  start to be  significant.
As an illustration,the muon decay  length is
\begin{equation}
L_\mu = \tau_\mu \;{p_\mu \over m_\mu} =  6.23~p_\mu({\rm GeV}) ~{\rm Km} 
\end{equation}
The  muons  are produced   at  an average   altitude of
$\sim 17.5$~Km   ($\sim 30$~Km) for  vertical
(horizontal)  particles  with only a weak  energy  dependence.
For  $p_\mu  < 1$~GeV,  most  muons  decay  independently  
from their direction.
For  $p_\mu = 3$~GeV  (10~GeV)    the decay of the   vertical
muons  is  suppressed by  a factor $\sim 2$ ($\sim 4$)
while the decay  probability  of horizontal  particles
remains  approximately  unity.

These  effects  are carefully included  in the  existing calculations,
that in fact   give     neutrino  zenith angle  distributions  that  are
approximately  flat   for  low $E_\nu$  (with   distortions  that  reflect
only  geomagnetic effects and
 the  angular distribution  of  the primary cosmic  ray flux),
 and    develop  a   stronger and stronger   enhancement on the horizontal
with increasing  energy.

\section {Spherical  geometry  of the source}

Most atmospheric  neutrinos  are produced  in  a  spherical shell
of  air at an altitude between  10 and 40 km.
The approximately spherical  geometry of  the  source region  has
some  very important effects  on the   angular distribution of the
neutrinos  and    results in   an enhancement of the  neutrino
 flux  from  the  horizontal  directions.
This  enhancement  has  been  overlooked in  all calculations
of the atmospheric  neutrino  fluxes before  the   work 
of Battistoni  et al.  \cite{fluka-3d}.

To  illustrate the nature of the new  effect,  let us  consider
a  situation where  the  geomagnetic  effects  are absent
(and therefore  the flux  of primary cosmic  rays  is exactly isotropic),
and the  zenith angle  dependence of the neutrino yield 
is  negligible,  that is  the number  of neutrinos
produced  by  a primary particle  does  not  depend
on  inclination of the shower in the atmosphere
(as  it is the case  for low  neutrino  energy).
In this  situation  a one--dimensional  calculation
predicts an isotropic  neutrino  flux. This  result is incorrect.
A realistic  (three--dimensional)
calculation  results   in an angular   distribution
of the neutrino flux  that    is   exactly  symmetric 
for  up--down    reflections, and  for   rotations  around the vertical axis
but that is {\em  not} isotropic and  exhibits  an
enhancement  for the  horizontal directions.
The enhancement for   the  horizontal  directions    depends on the neutrino
energy  and is more and more marked  with  decreasing energy.

The origin of  this result  can appear  surprising, and  for  
``pedagogical reasons'' it can  be  instructive to
consider  two   simple problems that  have clear  analogies
with the  emission  of  neutrinos in  thre atmosphere.
In problem A   an observer is  located  inside  a  thin spherical  shell
of    stars  of   radius $R_s$;  in problem  B  the  observer
is  inside  a spherical  cavity 
in a black body with temperature $T$.
In both cases we want to compute the 
angular  distribution of the
photon   flux measured  by the observer.

\subsection {Problem A: a shell of  stars  (isotropic emission)}
\begin{figure} [t]
\centerline{\psfig{figure=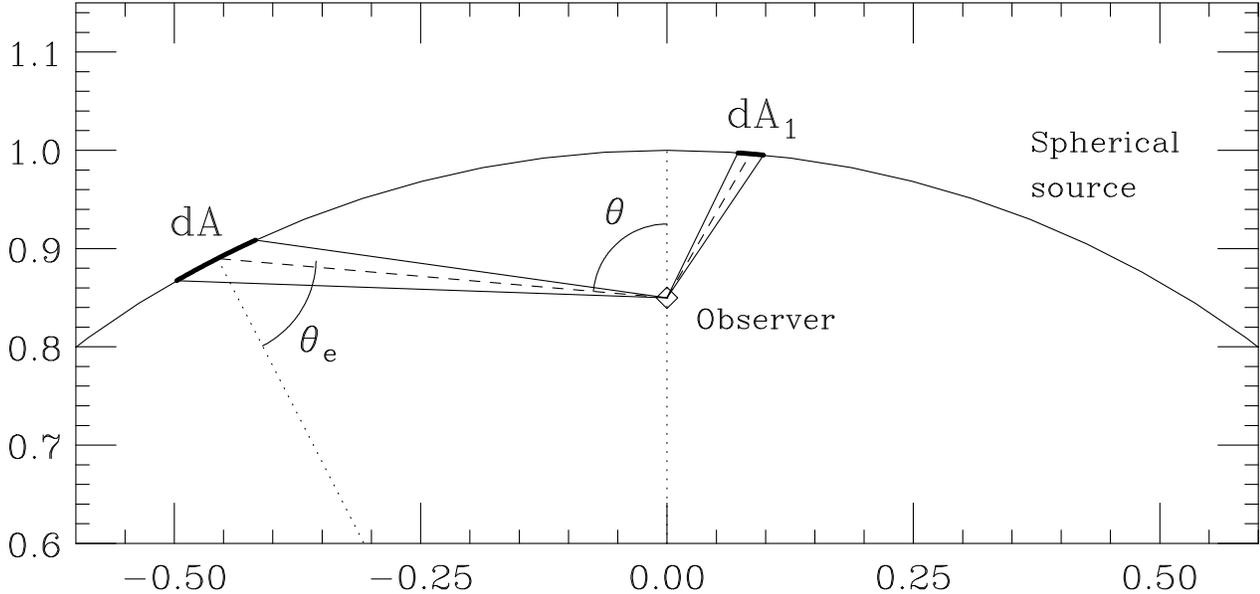,angle=90,height=8.0cm}}
\caption {Geometry of the   emission  from a thin
spherical  shell  with the observer  located  inside the shell.
\label{fig:3d-pedagogical}  }
\end{figure}

Let us  consider   an 
observer inside a  thin  shell  composed
of isotropically emitting stars (see fig.~\ref{fig:3d-pedagogical}). 
Let us assume  that 
 $L_0$ (erg/s) is the 
  average  luminosity  of the  stars,
$n$ is the surface number  density    (cm$^{-2}$) of stars  in the shell,
and  $R_A$ (with $R_A < R_s$)  the   distance of  the observer from the
center of the star  shell.
The   energy flux  per unit   solid  angle 
$dF(\Omega)/d\Omega$  
(erg/(cm$^2$~s~sr))  measured    by the observer can be calculated as:

\begin {equation}
\left ({dF(\Omega) \over d\Omega} \right )_{\rm iso}  
=   n ~{L_0  \over 4 \pi} ~
{1  \over \cos \theta_e} =
  n~{L_0  \over 4 \pi}
~\left [ 1  - \left( {R_A \over R_s} \right )^2 (1 - \cos^2 \theta)
 \right] ^{-{ 1 \over 2}}
\label{eq:3d-iso}
\end{equation}
where  $\theta$  (see  fig.~\ref{fig:3d-pedagogical}) 
 is  a polar  (``zenith'') angle around the axis
 that  passes  through the 
center of the  spherical   shell  and the observer.

\begin{figure} [t]
\centerline{\psfig{figure=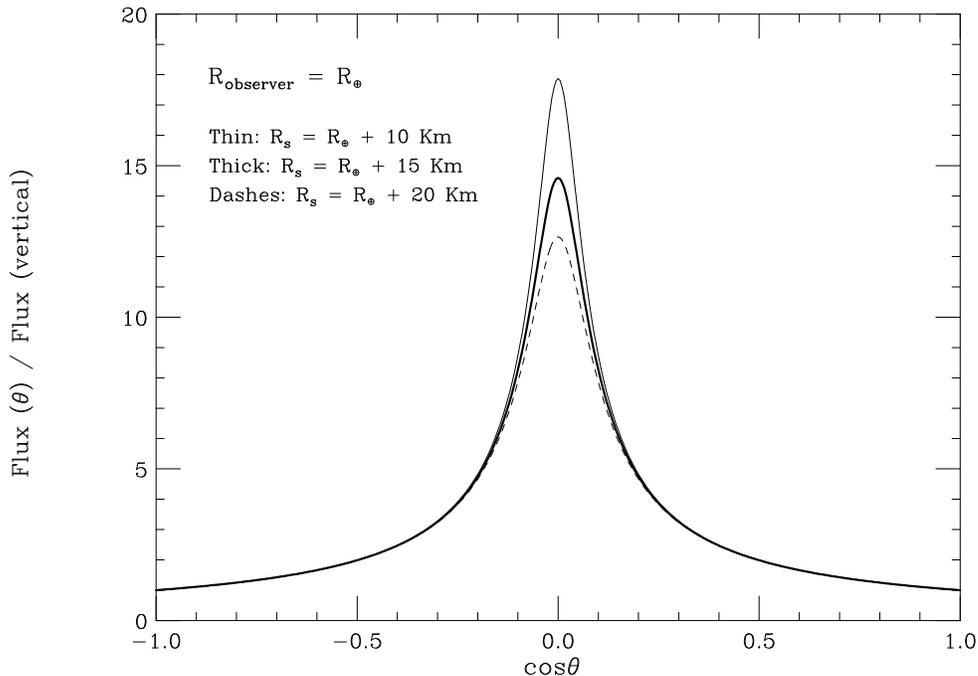,angle=90,height=9.0cm}}
\caption {Angular distribution of the flux  received 
by an  observer  located inside  a  shell of  isotropically
emitting material. $R_s$ is then radius of the emitting shell, 
and $R_{\rm observer}$ the distance of the observer from the center
of the shell. The  enhancement  for  $\theta \simeq 90^\circ$
(the horizontal plane)  becomes stronger when  $R_{\rm observer}$
approaches $R_s$. 
The  vertical  (up--going and down--going) fluxes  
are independent from $R_s$ and $R_{\rm observer}$.
\label{fig:iso}  }
\end{figure}

The solution  respects the 
cylindrical  symmetry  of our problem (does not
depend  on  the  azimuthal  angle  $\varphi$) and 
is  up--down symmetric (it remains
identical  after a reflection
$\theta \leftrightarrow \pi - \theta$), 
 but   neglecting the special
case of the  observer at the center of the shell,  it is 
{\em not}  spherically
symmetric, and  is  peaked  for horizontal  directions
($\cos \theta = 0$).
When the observer is  close to the  emitting shell
(that is when $R_A$  approaches $R_s$)
the peaking is   very strong,  and the   observer sees
the shell  as a   narrow  bright disk.
As a  numerical  example:  for  $R_A \simeq  R_\oplus  = 6371$~Km  
and  $R_s = R_\oplus + 15$~km, formula 
(\ref{eq:3d-iso})  predicts  a
horizontal   flux  14.6  times    more intense than the vertical 
flux (see fig.~\ref{fig:iso}).

It is  elementary to  deduce  equation (\ref{eq:3d-iso}).
Only stars  in the  surface element  $dA$  subtended   by the
solid  angle $d\Omega$ contribute to the    energy flux  
from  that  solid  angle interval  with a total of
$n~dA$ stars  (see  fig.~\ref{fig:3d-pedagogical});
each star   contributes an energy flux  $L_0/(4\pi \,\ell^2(\theta))$, 
where $\ell(\theta)$ is the distance  between the observer and the 
sources in the direction $\theta$.
Therefore:
\begin {equation}
 (dF)_{\rm iso} = n ~ dA ~ {L_0 \over 4 \pi \ell^2} = 
n~ { d\Omega \,  \ell^2 \over \cos \theta_e}~ {L_0 \over 4 \pi \ell^2}
\label{eq:3d-iso1}
\end{equation}
In the second  equality we have written  explicitely the expression
for  the area of the  element of shell  seen in  
the solid  angle $d\Omega$:  $dA = d\Omega \, \ell^2/\cos\theta_e$. 
The area  of the shell  element
 obviously scales  as  $\propto \ell^2$, and this  factor  
 cancels   exactly  the  factor $\ell^{-2}$
that takes into accout the
reduction with distance of the apparent  luminosity  of the 
stars, however   the expression for $dA$  contains also
a  geometrical  term 
$(\cos \theta_e)^{-1}$  that takes into  account the
orientation of the  surface element 
of the shell with respect to the line  of  sight.
The  ``emission'' angle $\theta_e$
 is the angle   between the
normal to the source  surface (toward the center of the shell)
and the line of  sight from the  source  to the  observer.
From elementary geometry one obtains:
\begin{equation}
\cos \theta_e = \sqrt{1 - 
\left ({ R_A \over R_s} \right)^2 (1 - \cos^2 \theta) }
\end{equation} 
It is  this   factor that is  responsible for   the strong 
enhancement  of the flux from the horizontal  direction.

It is interesting to  note that the   up--going and  down--going
   vertical fluxes  ($\cos \theta = \pm 1$) are 
independent from the position of the observer inside the shell,
while the  horizontal flux depends on the 
observer position, and    grows   when  the  observer approaches the 
emitting spherical shell. It follows that the total (angle 
integrated flux) is  a  function of the observer position:
\begin {equation}
F_{\rm tot}^{\rm iso}   = \int_{[4 \pi]} d\Omega~ 
\left ({dF(\Omega) \over d\Omega} \right )_{\rm iso} 
= n \, L_0  ~{1 \over  r} 
~\sinh^{-1} \left [ {r \over \sqrt{ 1-r^2} } \right ]
\end{equation}
(where  $r = R_A/R_s$)
and   grows  monotonically  from a value $n\,L_0$  for an observer
at the center of the  shell  to a divergent  value  when $r \to 1$.
The divergence is connected to the fact that
an  observer exactly  on  the shell must be (in the
approximation  we have used of a shell of vanishing  thickness) 
infinitesimally close to a star.

\subsection {Problem  B: a  cavity in a blackbody
 (emission $\propto\cos\theta_e$)}
The  solution of  problem B, is  of course
well kown,  we actually  live   in  a  (very large)  cavity whose
walls   have  a  temperature  of 2.7$^\circ$ Kelvin.
An  observer inside  a spherical    ``cavity''
in a black body (and in fact in a     cavity 
  of   arbitrary shape and dimension)
observes  an isotropic   black body spectrum  independently
from its position.
It can be  instructive  to  deduce  this  well known  result  from the same
considerations  used in the previous   discussion.
The surface  of a black  body  emits 
energy per  unit  surface and  unit solid  angle at a rate:
\begin{equation}
{ dL_{bb} \over d \Omega_e} =  \sigma\,T^4 ~{\cos \theta_e \over \pi}
\end{equation}
where $\sigma$ is the Stefan--Boltzmann constant.
The emission is  not   isotropic  but    decreases linearly with 
the  cosine of  the emission angle with  respect to the normal  to the 
surface element.  The  energy flux   from the direction $\Omega$
for the observer of problem  B   
can  be calculated  as:
\begin{equation}
 (dF)_{bb} =  dA~ {dL_{bb} \over d \Omega} ~ {1 \over \ell^2  } =
 \left (   { d\Omega \, \ell ^2\over \cos \theta_e} \right )
\left (  \sigma \,T^4 \; {\cos \theta_e \over \pi} \right )
~
\left ( { 1 \over \ell^2 } \right )
= {\sigma \,T^4 \over \pi} ~ d\Omega
\label{eq:3d-bb}
\end{equation}
Note that in this case   there is  a cancellation  not only for 
the factor   $\ell^2$, but also for the 
orientation  factor $\cos \theta_e$. 
If  the normal to emitting surface  has a
large  angle with the line  of   sight, a larger surface  ($\propto
(\cos \theta_e)^{-1}$)
can  contribute to the emission, but the
 large angle emission  is suppressed by a  factor
$\cos \theta_e$.  Combining  the two effects one has  
  an  exact cancellation.
The demonstration has  been actually  general, and is valid
for  a cavity of arbitrary shape or dimension  and for 
an arbitrary position of the observer.

\subsection {Atmospheric neutrinos}
We can  now  come back to  the problem of  atmospheric  neutrinos.
The source  volume of atmospheric  neutrinos
is  also  with  very good approximation a spherical  shell
with the  observers  placed inside,  very close to the inner radius  of the 
source  volume.
However an  element of  atmosphere  emits  neutrinos  with
an angular  distribution that is  neither  isotropic,  nor
linear in  zenith angle, therefore  we  can  expect a 
neutrino   flux  at  sea  level with an  angular  distribution
with a form that is  intermediate   between  
the solution for  problem A  (isotropic emission in the source
and  strong  horizontal  peaking of the observed flux)
and  problem B (emission $\propto  \cos \theta_e$   in the source
and isotropic  observed  flux).

Let us consider
an isotropic flux  of cosmic   rays  that   impinges on the Earth atmosphere.
The   quantity  $C_0$    that is 
the  number of   cosmic  rays   absorbed   per unit time 
by an element  of unit area of the atmosphere   (units cm$^{-2}$~s$^{-1}$)   
can  be  calculated as:
\begin {equation}
C_0 = \int_{[2 \pi]} d\Omega ~\phi_0 \, \cos \theta_0  = \pi \, \phi_0
\label{eq:prim-abs}
\end{equation}
(the integration is  over  one hemisphere). 
Note that  the absorption rate of
cosmic rays is proportional  to  $\cos \theta_0$,  with
$\theta_0$ the angle between the normal to the surface element
and the cosmic  ray   direction.
If the cosmic  rays  produce  an average number $\langle n_\nu\rangle$
of neutrinos   per  showers, then each element of   the atmosphere is
also  a source of neutrinos, with an emission rate
$S_\nu   = C_0\, \langle n_\nu \rangle$   (again in units cm$^{-2}$~s$^{-1}$).

The angular  distribution  
of  the  neutrino  emission
from  a  ``patch'' of  atmosphere  is  not easy to calculate  and
depends   on  the average angle  between  neutrino
and  primary  particle.
Knowing this  distribution  it is  possible to 
calculate  the 
observable flux  at sea level,   following the same  steps
outlined  above.

The  results for two limiting cases  are easy  to obtain.
If the neutrinos  in a shower  are produced   collinearly with the primary
trajectory ($\Omega_\nu = \Omega_0$) 
then the  angular  distribution of the neutrino emission
from a  surface element  of  the atmosphere  is 
$\propto \cos \theta_\nu$, because  it
simply  reflects  the  angular  distribution   of the 
absorbed primary flux  (see equation (\ref{eq:prim-abs})).
From equation (\ref{eq:3d-bb})
(neglecting  the   angular dependence  of the neutrino  yield)
it  follows  that  the observed 
neutrino flux is  isotropic.

%If the  neutrino yield  
%depends on the absorption angle of the primary particle
%the neutrino flux is
%$\phi_\nu(theta) = \phi_0 \,\langle n_\nu (\theta) \rangle$
%(where $\theta$ is the zenith  angle of  both the neutrino and primary
%parent)  and  the  angular distribution only reflects the 
%angular dependence of the yield.

The opposite  limiting case is   obtained  when then
neutrinos  are  emitted  quasi isotropically
in the primary cosmic  ray  showers.
This  is    approximately true  only  for  very low neutrino  energy, 
($E_\nu \aprle 10$~MeV).
In this  case  all  memory of the  
direction of the primaries is  erased,  and the neutrino emission from
an element of  the atmosphere is  isotropic.
From equation (\ref{eq:3d-iso}) it follows that the angular  distribution of
the observed neutrino  fluxes is  very sharply  
peaked  for the  horizontal directions.
Considering the altitude distribution of the neutrino 
production points
(that is  related  to the  density  profile of the atmosphere
and the value of the hadronic  cross sections) 
it can be estimated 
that for  the   very low energy neutrinos that are emitted
 quasi--isotropically, the  flux  
on the horizontal plane
is more  than an order of magnitude
more intense than in the vertical  directions (see fig.~\ref{fig:iso}).

In  general neutrinos  are emitted in a cone
with an axis  that corresponds to  the primary particle direction
and an  opening  angle that shrinks with increasing   neutrino
energy.  We can therefore expect that 
the  emission of  the neutrinos from the atmosphere is
in  general  intermediate, between  the
two  extremes   corresponding  to  isotropic  emission, and  an emission
$\propto  \cos \theta_e$.
The two  limiting cases  are  approached for  very low
neutrino  energy, when the  neutrino emission is  very poorly correlated
to the primary  direction,  and for high energy neutrinos
when   because  of the Lorentz boost, the neutrinos  are emitted
quasi--collinearly with the primary particle.

\begin{figure} [t]
\centerline{\psfig{figure=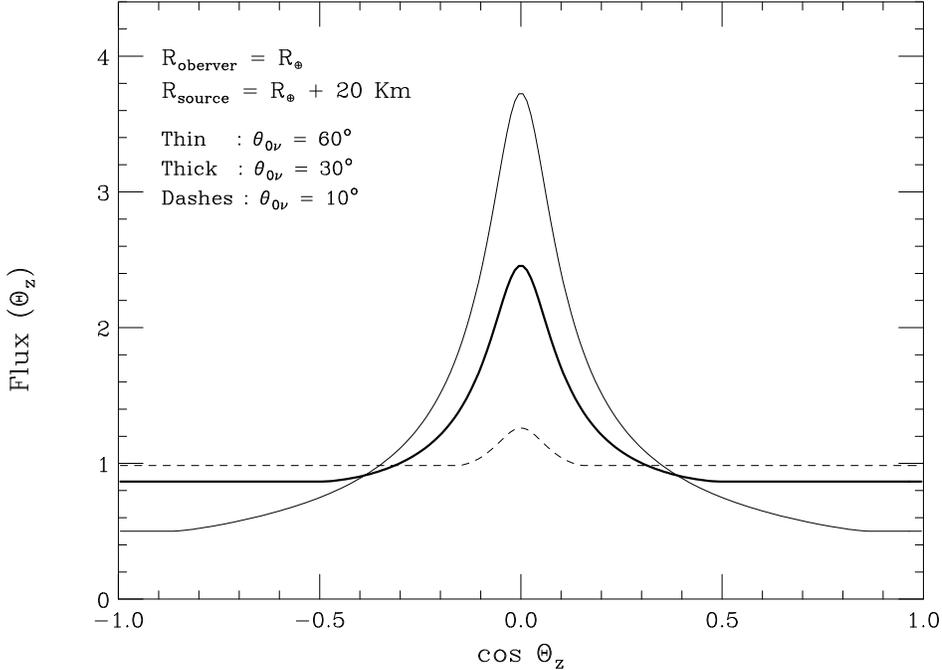,angle=90,height=9.0cm}}

\caption {Angular  distributions 
 of  the neutrino  flux 
emitted by   a   spherical thin layer
of atmosphere of radius $R_s = R_\oplus  + 20$~Km, and observed 
at sea level. 
It is  assumed  that the  atmosphere layer   receives
an isotropic  flux   of  cosmic  rays, that the average 
number of neutrinos  produced in a shower is  independent from 
its zenith angle, and that the
angle between the $\nu$ and the primary particle has a fixed value.
The three curves  are calculated  for  values
$\theta_{0\nu} = 10^\circ$, $30^\circ$ and  60$^\circ$.
With  increasing $\theta_{0\nu}$ the 
vertical flux  becomes  more and more suppressed  and
the  horizontal flux  more and  more enhanced.
\label{fig:alpha1}
}
\end{figure}

\begin{figure} [t]
\centerline{\psfig{figure=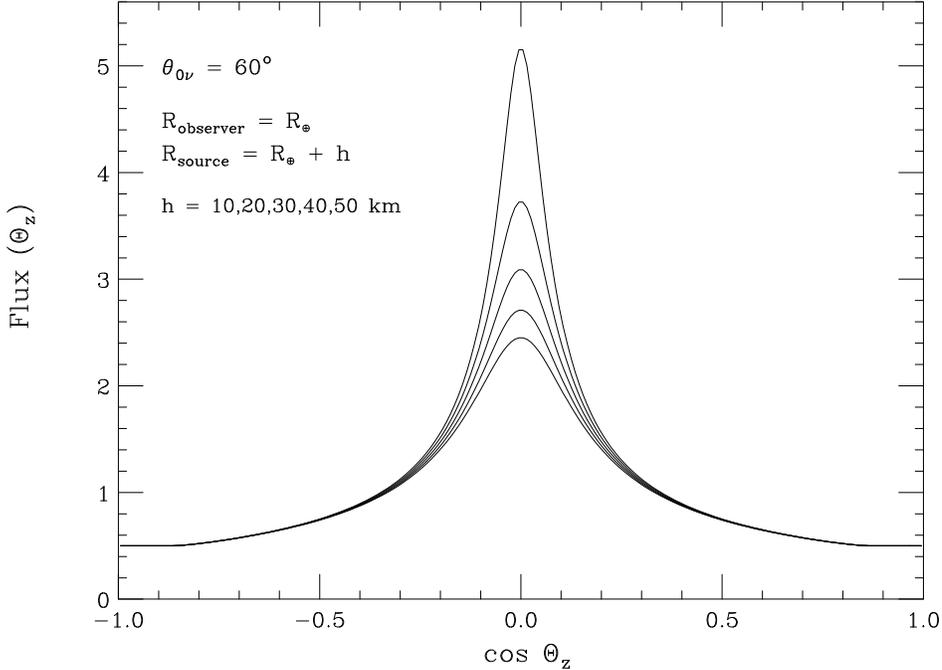,angle=90,height=9.0cm}}

\caption {Angular  distributions 
 of the neutrinos  produced  by a  thin  spherical layer
   of atmosphere of radius $R_s = R_\oplus + h$ 
with $h =  10$, 20, 30, 40 and 50~Km.
It is assumed  that the   layer of atmosphere  absorbs an isotropic
flux  of cosmic rays  and that the
neutrinos  are  emitted  at a fixed  angle $\theta_{0\nu} = 60^\circ$
with respect  to the   primary particle direction. 
The  enhancement for the horizontal  directions
becomes  stronger when the  emission layer  is  closer
to the observer.  The vertical  (up--going or  down--going)  flux
is  independent from the  radius  of the emitting layer.
\label{fig:alpha2}
}
\end{figure}

These expectations  can be  verified with more  detailed  calculations.
As an illustration, let us  consider  the  simplified problem
where   (i) the  atmosphere  is    formed  by a single thin  layer
at  an height $h$ (the  radius of the emitting layer 
is  therefore $R_s = R_\oplus + h$); (ii)   the  primary cosmic  ray  flux
is  exactly  isotropic; (iii) the neutrino  production    is  independent
of the zenith  angle of the primary particle; and (iv) 
the  angle  $\theta_{0\nu}$ between each  neutrino  and  its primary particle
has  a fixed  value  $\theta_{0\nu} = \alpha$
(and the emission  has cylindrical  symmetry around the  axis
defined by the primary particle trajectory).
The problem  is to compute the 
the   neutrino  flux received   from an  observer located at sea level.
This   simplified problem  contains  essentially all the  interesting
geometry of atmospheric  neutrino  production \cite{fluka-3d}.
Defining  $r = R_\oplus/R_s$, and
 $S_\nu = \pi \phi_0 \langle n_\nu\rangle$ the  number  of  
 neutrinos  emitted per unit time  and unit surface  by the emitting layer,
the solution   can  be written as:
\begin{equation}
\phi_\nu (\Omega_\nu, r, \alpha) =  {S_\nu \over 2 \pi}
~\left  [ {F_\alpha(y) \over  y} 
 \right ]_{y = \sqrt{1- r^2 (1-\cos^2 \theta_\nu)}}
\label{eq:spherical}
\end{equation}
with  (for $\alpha < 90^\circ$):
{\normalsize
\begin{equation}
F_\alpha(y) = \cases { 
0          &for $y < - \sin \alpha$,  \cr
\omit      & \omit \cr
\cos \alpha \; y + {2 \over \pi} \sqrt{1 - \cos^2 \alpha - y^2} 
- {2 \over \pi} \cos \alpha \; y ~ \tan^{-1} \left [
{ \cos \alpha \; y \; \sqrt{1 - \cos^2 \alpha - y^2}  \over 
\cos^2 \alpha + y^2 -1 } \right ]
  & for $y < |\sin \alpha|$ , \cr
\omit      & \omit \cr
2\, \cos \alpha \,y  &for $y  > \sin \alpha$.  \cr
 }
\label{eq:spherical1}
\end{equation} 
}
For $\alpha > 90^\circ$  one has  to  substitute:
$F_{\alpha}(y)  = F_{\pi -\alpha} (-y)$.
This  solution  can be  easily  checked  numerically with  a montecarlo
method writing a simple program  with few lines of computer code.
Some  examples of the solution  are shown in fig.~\ref{fig:alpha1}
and fig.~\ref{fig:alpha2}.
For  $\theta_{0\nu} \to 0$  the     observed neutrino flux
is  isotropic  with  a  value
 $\phi_\nu = \phi_0 \,\langle n_\nu \rangle$
where $\phi_0$ is the isotropic  primary flux
and $\langle n_\nu \rangle$  the average number of neutrino
produced   by a  primary  particle.
When  the  angle  $\theta_{0\nu}$ grows the   observed  flux
is  suppressed  for the vertical  direction
and is enhanced   on the horizontal plane. 
The  suppression and  enhancement  
become more important
when the  angle $\theta_{0\nu}$ increases.
For  example (see fig.~\ref{fig:alpha1})
   if the   neutrinos  are produced at an
height  $h = 20$~km  and 
are emitted  with an  angle $\theta_{0\nu} =10^\circ$, 
 $30^\circ$, or $60^\circ$  with  respect 
to the primary particle direction,  the observed   neutrino  flux
at  sea level  has  a horizontal/vertical ratio of 1.45, 3.25, and 8.5.

It is  important to observe  that the enhancement for the 
horizontal  directions    for  a fixed angle 
$\theta_{0\nu}$  depends  on the  ratio  $r = R_\oplus/R_s$
(that is on the  altitude  of the neutrino  production),  and
is  stronger when   the altitude  decreases.
On the contrary the suppression for the vertical  direction
does  depend on the angle $\theta_{0\nu}$ but  is independent of
the ratio  $R_\oplus/R_s$ (see equation (\ref{eq:spherical1})
and  fig.~\ref{fig:alpha2}).
An  important  consequence of this is  the fact   that  the average 
(angle  integrated)  neutrino  flux  observed at sea level
depends  on the altitude of  neutrino  production, and grows  when 
the  average  altitude  of production  decreases.
This can  easily understood qualitatively  observing that  for 
finite $\theta_{0\nu}$ a  fraction of the  produced  neutrinos
will ``miss'' the Earth    because are emitted  on  trajectories
that  do not intersect her  surface. The  fraction  neutrinos
``lost in space''
will  decrease  if the  production points  are   closer to the  surface.

It is  less intuitive to  note  that 
 for  large
$\theta_{0\nu}$  the flux
of neutrinos  in a full 3--D calculations  can be {\em higher}
than what is  obtained in a collinear  approximation.
This can  be checked   integrating  over  angle  equation
(\ref{eq:spherical}) and it is  possible
 because  the   enhancement on  the horizontal plane
can  be stronger   than  the suppression  on the vertical.
This  does not   represent a violation of  ``unitarity''
since it can   happen  only for a small range of  radii
just below the source  region in the atmosphere, 
  while  in most of the volume inside the Earth 
for larger $\theta_{0\nu}$ the flux  is suppressed.

From fig.~\ref{fig:iso} and fig.~\ref{fig:alpha2} one can see that
the  height of  neutrino  production is  very important in
determining the  enhancement for the horizontal directions.
To compute  the  geometrical effects for  atmospheric neutrinos
it is  not possible 
to  approximate  the emission volume
as  a thin surface.
The results for  a thick  shell  can of course be obtained
integrating   over a continuos  distributions  
of  thin shells with different radii,
The assumption of emission from an altitude 
of $\sim 15$--20~km   can  also give  a reasonable first order approximation.

In summary  this ``geometrical effect''   can be
very important  for low  neutrino  energy.
In a realistic  calculation it must be of course be 
combined  with the other 
effects: geomagnetic  effects
and the  zenith angle 
dependence of the neutrino  yields
to  determine the angular  distribution of  atmospheric  neutrinos.

It has  been  stated that the  enhancement on the horizontal 
is the result of the  inclusion   in a 3--D calculation of
primary particles   with   ``grazing''  trajectories  
that if continued would  not intersect the Earth's  surface
and are not  included  in a 1--D calculation.
These  grazing  trajectories  give a neglible  contribution to
the  horizontal neutrino fluxes, and are not the source of the
enhancement.
The   large  flux  from the horizontal directions
can  be   understood as the effect of  
the production of  ``horizontal''  neutrinos  from the 
more  numerous ``vertical''  showers after  emission
with an  appropriate angle  with respect to the shower axis.
Note than  an  isotropic  flux  implies that most of the
interacting tracks are in fact   vertical 
(the absorption rate is $\propto \cos \theta_0$), and 
therefore  the    opposite effects  of ``vertical''   showers producing 
``horizontal'' neutrinos  and  vice versa are  of different
size and do {\em not} cancel  each  other.
Note also  that the 
 enhancement for  the  horizontal directions is  linked
with  a suppression of the  vertical  fluxes.

A second  3--D calculation of the  atmospheric  neutrino fluxes
(Tserkovnyak  et al. \cite{3d-sno})  has been  recently made  public.
In this  calculation the     enhancement  of   the neutrino  fluxes
for the horizontal direction is not  present,  and the predicted
angular distributions  of the neutrinos
 are  very similar to those  calculated
in  a 1--D approach. 
The reasons  for  the  discrepancy 
with the Battistoni et al work \cite{fluka-3d}   are  not 
easily  understandable from a simple  reading of the papers.
Also  an  early attempt at a  3--D calculation 
of Lee and  Koh \cite{LK}   did not find a  flux enhancement 
for horizontal  directions.
As we are trying to illustrate
in this  work  the  enhancement  of the neutrino  fluxes
in  the  horizontal directions   is  the consequence of
 simple  geometry and {\em  must} be present
in a correctly performed   3--D calculation.   The absence of
this  effect in    \cite{3d-sno} and \cite{LK}  is therefore  evidence
of the   existence  of errors it the calculation method used
for the prediction.

\section {A complete   three-dimensional calculation}

To illustrate the points discussed in the previous  section
with a concrete  example we have performed  a  detailed  
calculation of the neutrino  fluxes with a three--dimensional method.

The method  of the calculation is  conceptually very simple.
The space  around the Earth  has  been  divided into 
two  regions:  an outer  region  
($r > R_\oplus + \Delta R$  where  $R_\oplus = 6371.2$~Km is the
Earth's average radius  and
$\Delta R =  80$~km)   where   the  matter  density
is  assumed  to be vanishingly small and
only   a magnetic  field is  present,  and an inner region
($R_\oplus < r < R_\oplus + \Delta R$)
where  particles  interact  not only
with the magnetic  field 
but  also with  air. The air is  modeled  as  having a
spherically  symmetric  density   profile $\rho(r)$.
The value  of $\Delta R$  was    determined  calculating
that less that one percent of  the  cosmic  rays that 
graze the  Earth  with a  distance  of  closest  approach  
equal to  $\Delta R$  interact in the   residual  
high  altitude atmosphere.
The result of the calculation are  independent  of
$\Delta R$  (if it is sufficiently large), but the calculation
become  inefficient for  $\Delta R$  too large,
 
The   fluxes  of cosmic   rays    entering the spherical surface 
at $R_s = R_\oplus + \Delta R$   were   determined  
using  a model   for the  isotropic  flux 
in the absence of geomagnetic  effects  (we used the results of
primary flux of the Bartol calculation \cite{bartol}),
and   taking into  account the  effects 
of the geomagnetic  field  (we used the IGRF field
model  \cite {igrf})  in the propagation
of the particles.
This   was  obtained    generating a   uniform and isotropic
flux  that  enters the spherical surface $R_s$, and rejecting
the particles  that  correspond to  forbidden  trajectories.

The primary  particles  are then propagated  inside the  inner  shell, 
taking into account  the  bending in the magnetic  field
and the interactions  with  the air nuclei.
When  an inelastic  interaction occurs,  a  set of  final state particles
is generated with a  realistic  distributions, of  multiplicity,
flavor,  energy  and   transverse momentum.
In this  calculation we have used a simple  model of the
hadronic  interactions  due to Hillas \cite{Hillas}. 
A  fraction of  approximately 3\% of the primary  cosmic  rays that enter 
the  inner shell  exit  without interacting.

The  subsequent  development  of the shower is  a standard one.
The trajectories of
charged   particles are  propagated  along
curved   trajectories  because of the  presence of the magnetic
field.  When  neutrinos  are produced their trajectories
(simple  straight lines)  are  studied.  The  trajectories  intersect
$n$ times  the  surface of the Earth with $n=0$ or
$n=2$.  If $n=0$  the neutrino is  discarded. If $n=2$  two  neutrinos
(corresponding to down--going and up--going trajectories)
are ``collected''  and  their  position and direction are recorded.

\begin{figure} [t]
\centerline{\psfig{figure=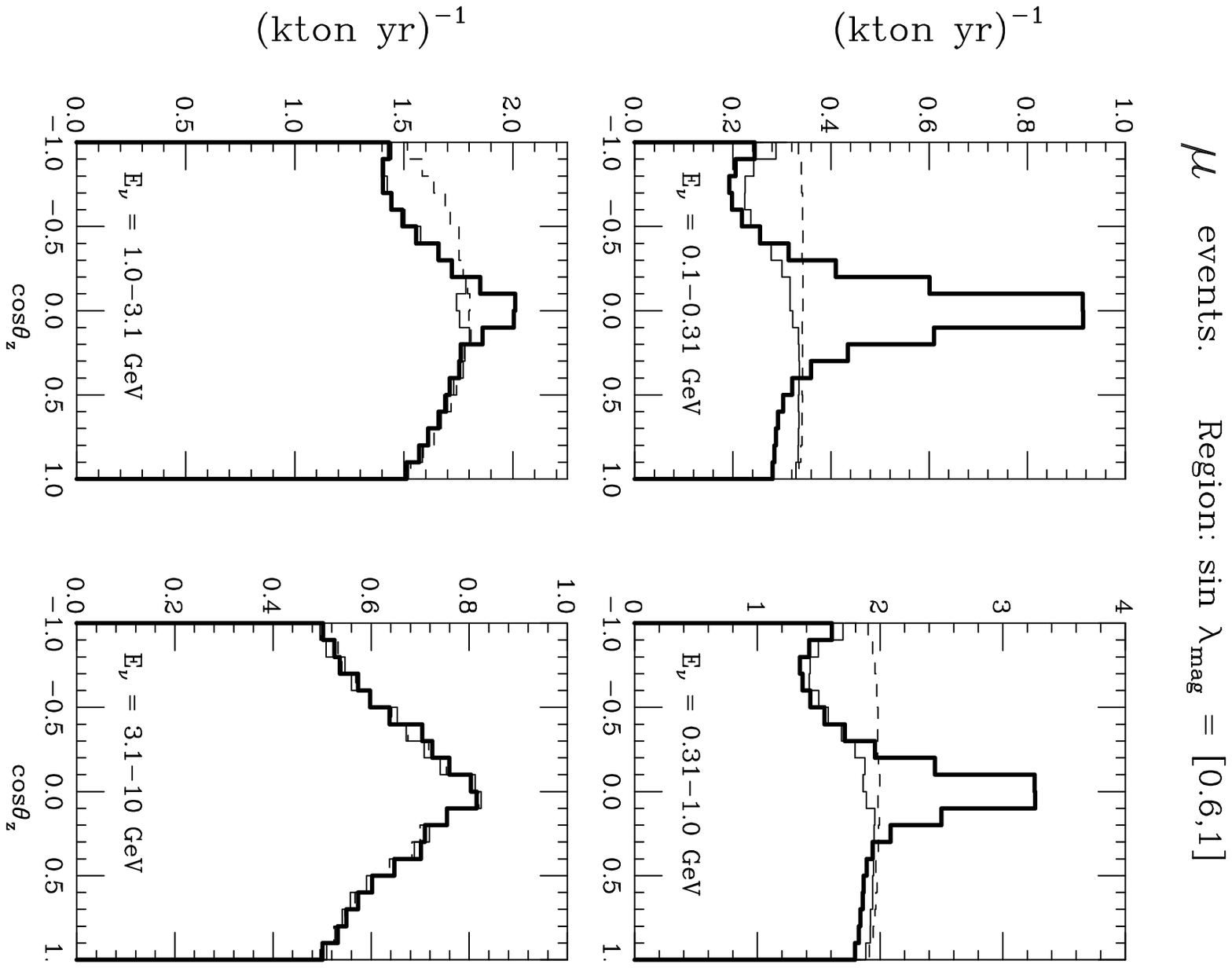,angle=90,height=18.0cm}}
\caption {Average zenith angle  distributions  for $\mu$--like events 
for  detectors located  in positions on the Earth with magnetic  latitude
$ \sin \lambda_M > 0.6$ (magnetic  north polar region).
The four panel  are for   different  neutrino energy intervals.
The  three histograms  are  for: fully 3--D calculation
(thick), 1--D calculation (thin),  1--D without geomagnetic  effects
(dashed).  
\label{fig:mu_1}  }
\end{figure}

\begin{figure} [t]
\centerline{\psfig{figure=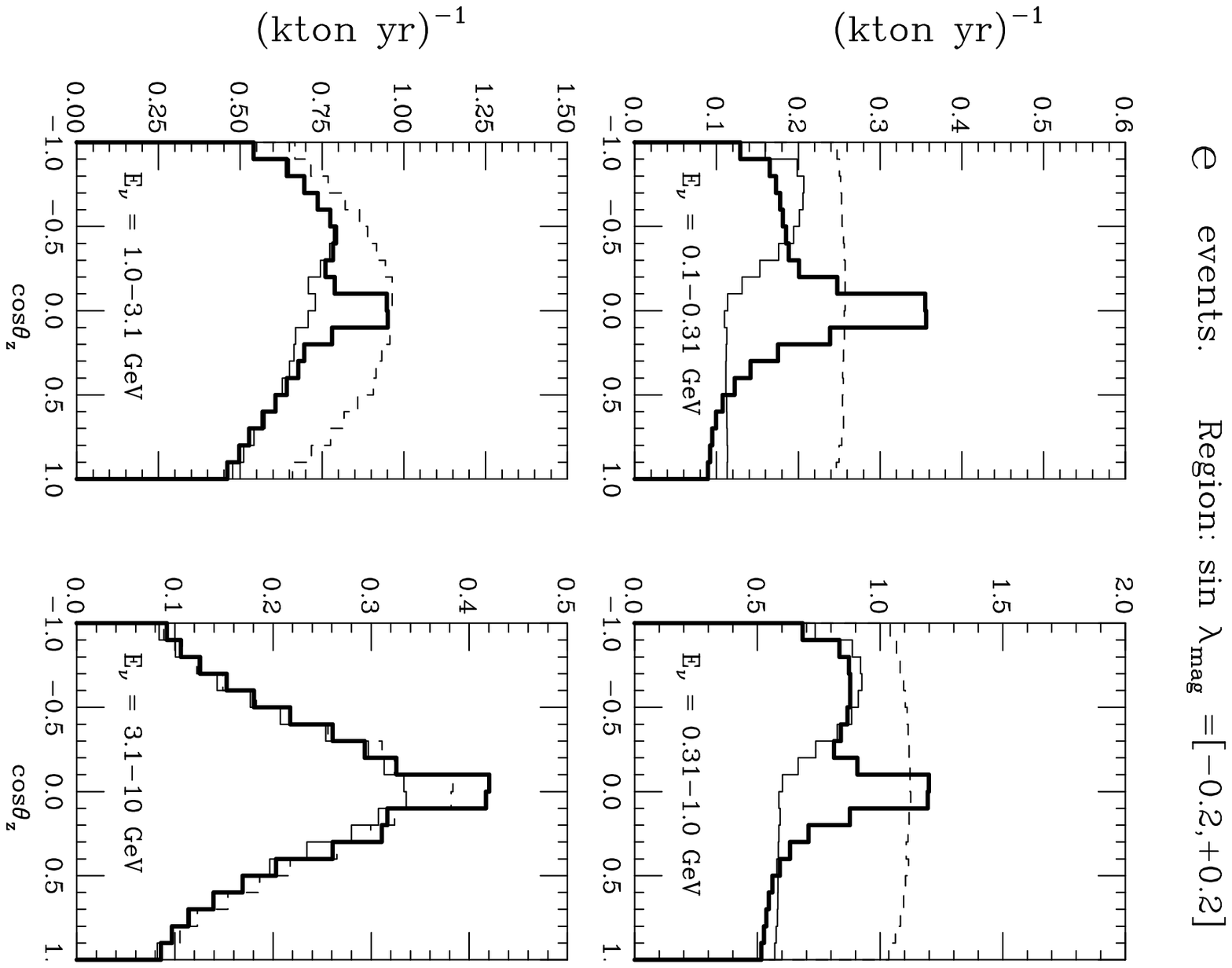,angle=90,height=18.0cm}}
\caption {Average zenith angle  distributions  for $e$--like events 
for  detectors located  in positions on the Earth with magnetic  latitude
$ \sin \lambda_M = [0.2,0.6]$.
The four panel  are for   different  neutrino energy intervals.
The  three histograms  are  for: fully 3--D calculation
(thick), 1--D calculation (thin),  1--D without geomagnetic  effects
(dashed).  
\label{fig:e_3}  }
\end{figure}

Some of the  results  of our  calculation  are collected
in six figures  (from fig.~\ref{fig:mu_1}  to fig.~\ref{fig:e_3}).
For illustration (and debugging purposes) we have 
performed  the calculation three times.

\begin{itemize}
\item [(i)] A first  calculation  (represented by the thick) histograms
was  performed  using  a fully 3--D method. 
This includes   geomagnetic  effects on the  primary cosmic ray fluxes, 
and   in the shower development
the inclusion of   realistic $p_\perp$  distributions  in hadronic
interactions,  and the correct  treatment of the kinematics in particle
decays.

\item [(ii)] A second  calculation (represented by thin histograms)
was   performed  to   reproduce  the  1--D  algorithms.  
The geomagnetic  effects  are calculated  for  the primary cosmic
ray particles (exactly as in the previous case)
but  not for  the shower development
(particles  travel along straight lines for $r < R_\oplus + 80$~Km.),
and all   final  state particles  are collinear  with the projectile
(or  parent)  particle.  This  is achieved 
modeling    the interactions  and     particle  decays 
exactly as  in the previous case,
(including therefore transverse momentum), 
and  performing as a last step a rotation  of all 
the 3--momenta  of the final  state particles 
so that  they  become   parallel  to the projectile
(for  interactions)  or  parent (for decays) particle.

\item [(iii)] A third  calculation  (represented  by  dashed  histograms)
was performed  neglecting
the  geomagnetic effects on the primary flux (therefore  considering
exactly isotropic  primary fluxes)  and  using the 1--D
algorithms  outlined in the previous point.
\end {itemize}

Since the  atmospheric neutrino  fluxes  depend
 on the geographycal position of the detector,  we have  
divided  the Earth's  surfaces  into  five  equal area  regions, 
according to the  geomagnetic  latitude:
two polar regions, two   intermediate  latitude regions  and
an equatorial region, and have  collected  separately
 all  neutrinos  generated
in the montecarlo  calculation  that land in the different regions
(each neutrino is  recorded   in two separate  regions,
once as   down--going  and once as up--going).
The results in the two (north and south)  polar regions and
the  two  intermediate latitude regions are essentially   undistinguishable,
and therefore only the  results for the north regions are presented  here.

\begin{figure} [t]
\centerline{\psfig{figure=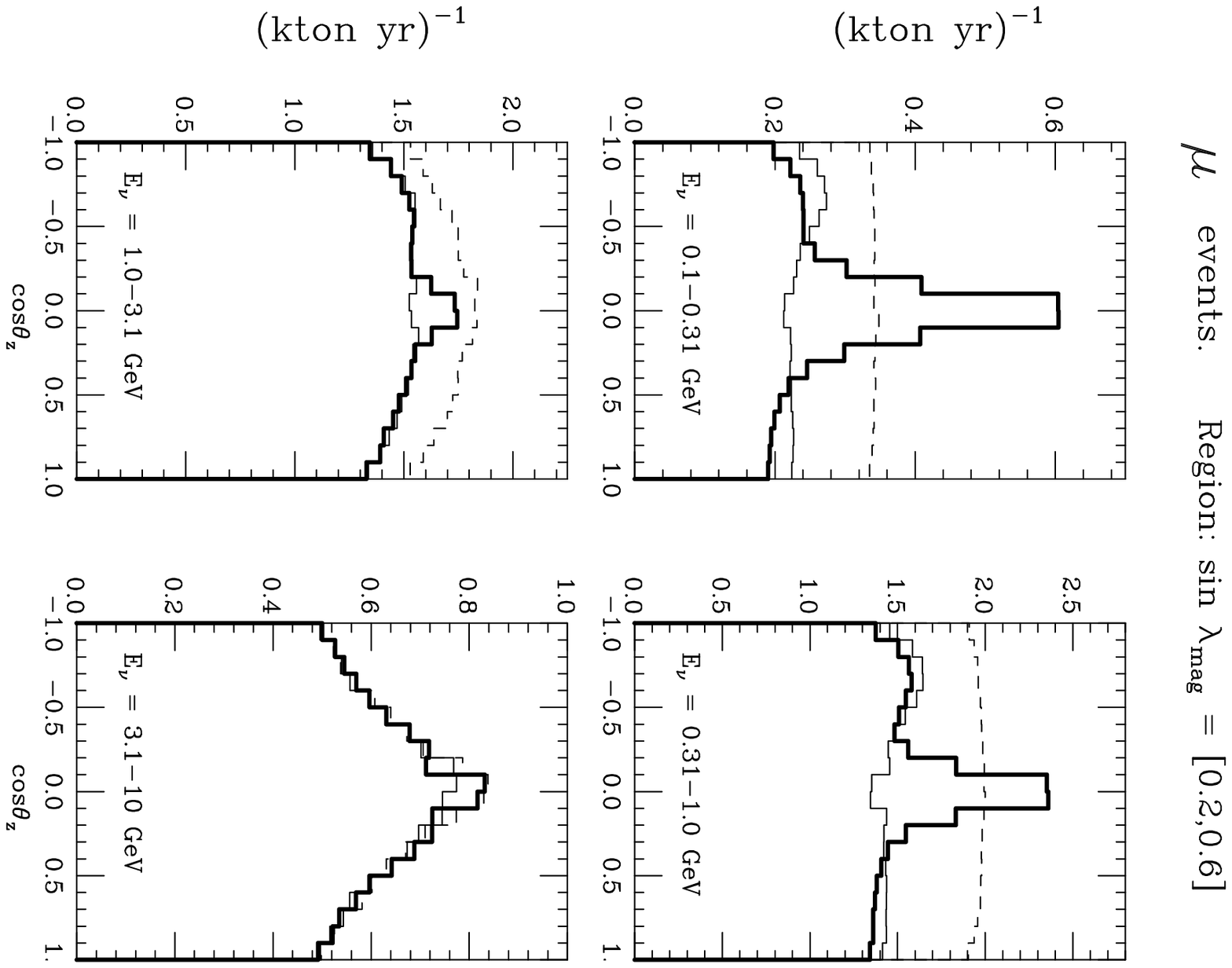,angle=90,height=18.0cm}}
\caption {Average zenith angle  distributions  for $\mu$--like events 
for  detectors located  in positions on the Earth with magnetic  latitude
$ \sin \lambda_M = [0.2,0.6]$.
The four panel  are for   different  neutrino energy intervals.
The  three histograms  are  for: fully 3--D calculation
(thick), 1--D calculation (thin),  1--D without geomagnetic  effects
(dashed).  
\label{fig:mu_2}  }
\end{figure}

\begin{figure} [t]
\centerline{\psfig{figure=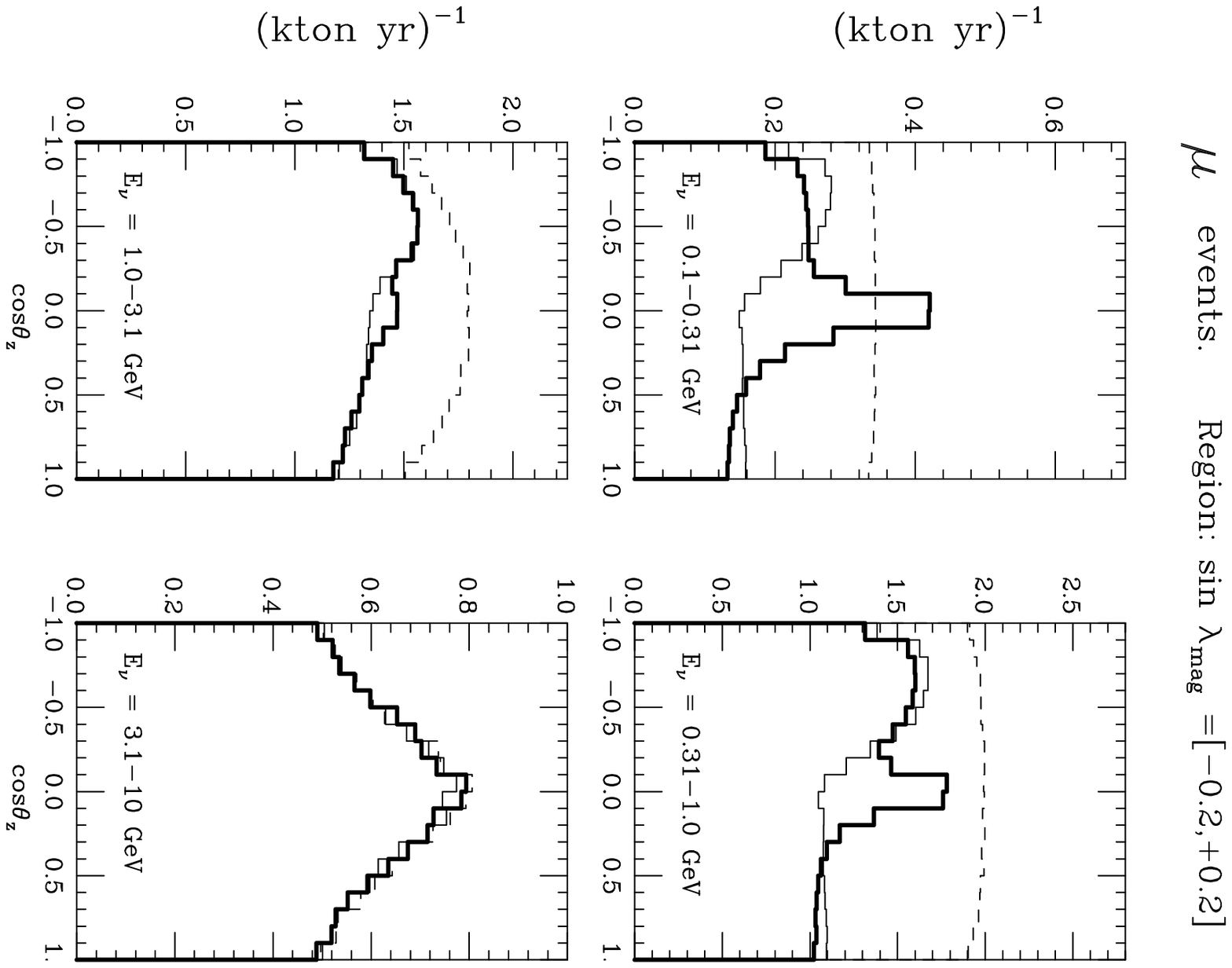,angle=90,height=18.0cm}}
\caption {Average zenith angle  distributions  for $\mu$--like events 
for  detectors located  in positions on the Earth with magnetic  latitude
$ \sin \lambda_M = [-0.2,+0.2]$ (magnetic  equatorial  region).
The four panel  are for   different  neutrino energy intervals.
The  three histograms  are  for: fully 3--D calculation
(thick), 1--D calculation (thin),  1--D without geomagnetic  effects
(dashed).  
\label{fig:mu_3}  }
\end{figure}

\begin{figure} [t]
\centerline{\psfig{figure=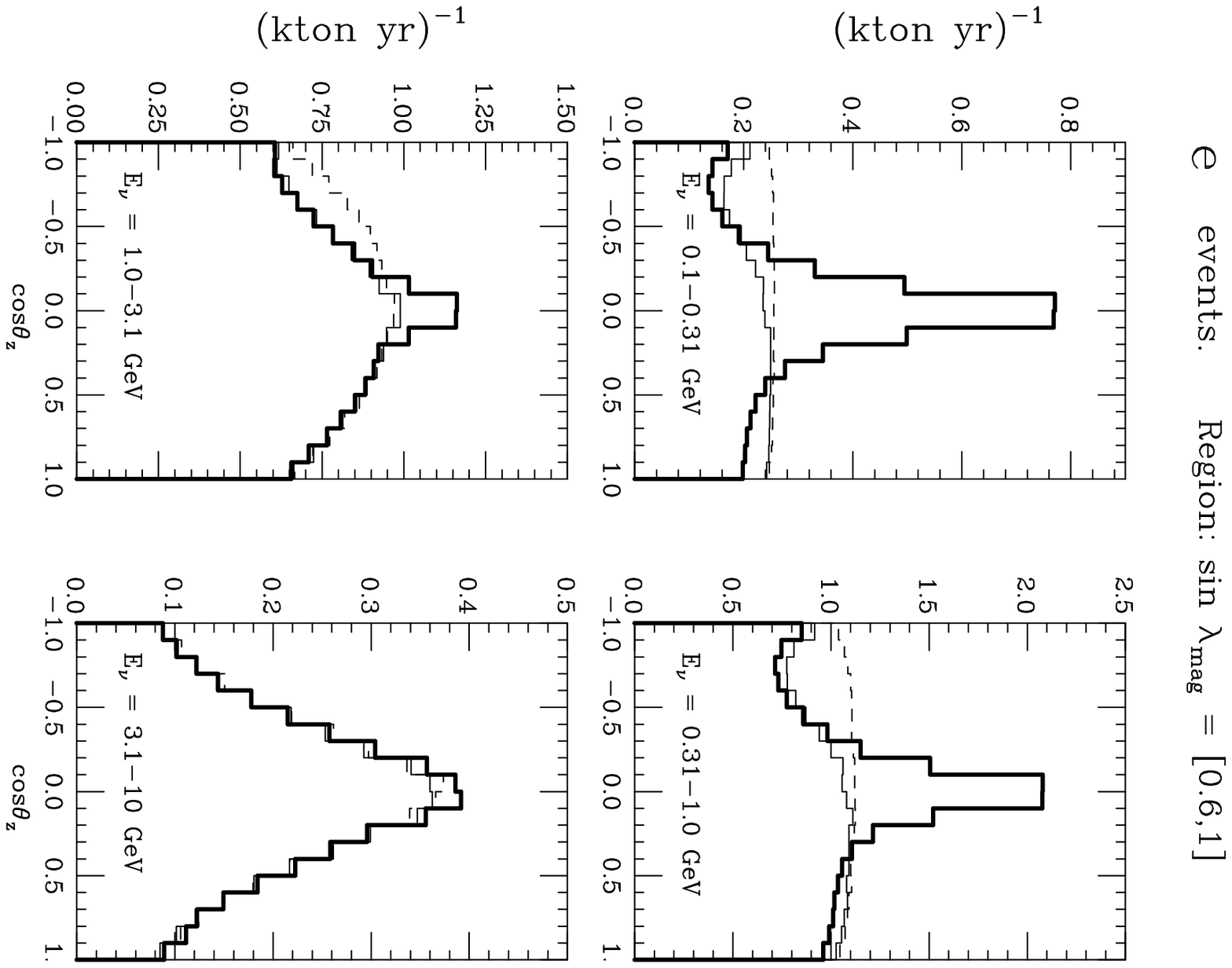,angle=90,height=18.0cm}}
\caption {Average zenith angle  distributions  for $e$--like events 
for  detectors located  in positions on the Earth with magnetic  latitude
$ \sin \lambda_M > 0.6$ (magnetic  north polar region).
The four panel  are for   different  neutrino energy intervals.
The  three histograms  are  for: fully 3--D calculation
(thick), 1--D calculation (thin),  1--D without geomagnetic  effects
(dashed).  
\label{fig:e_1}  }
\end{figure}

\begin{figure} [t]
\centerline{\psfig{figure=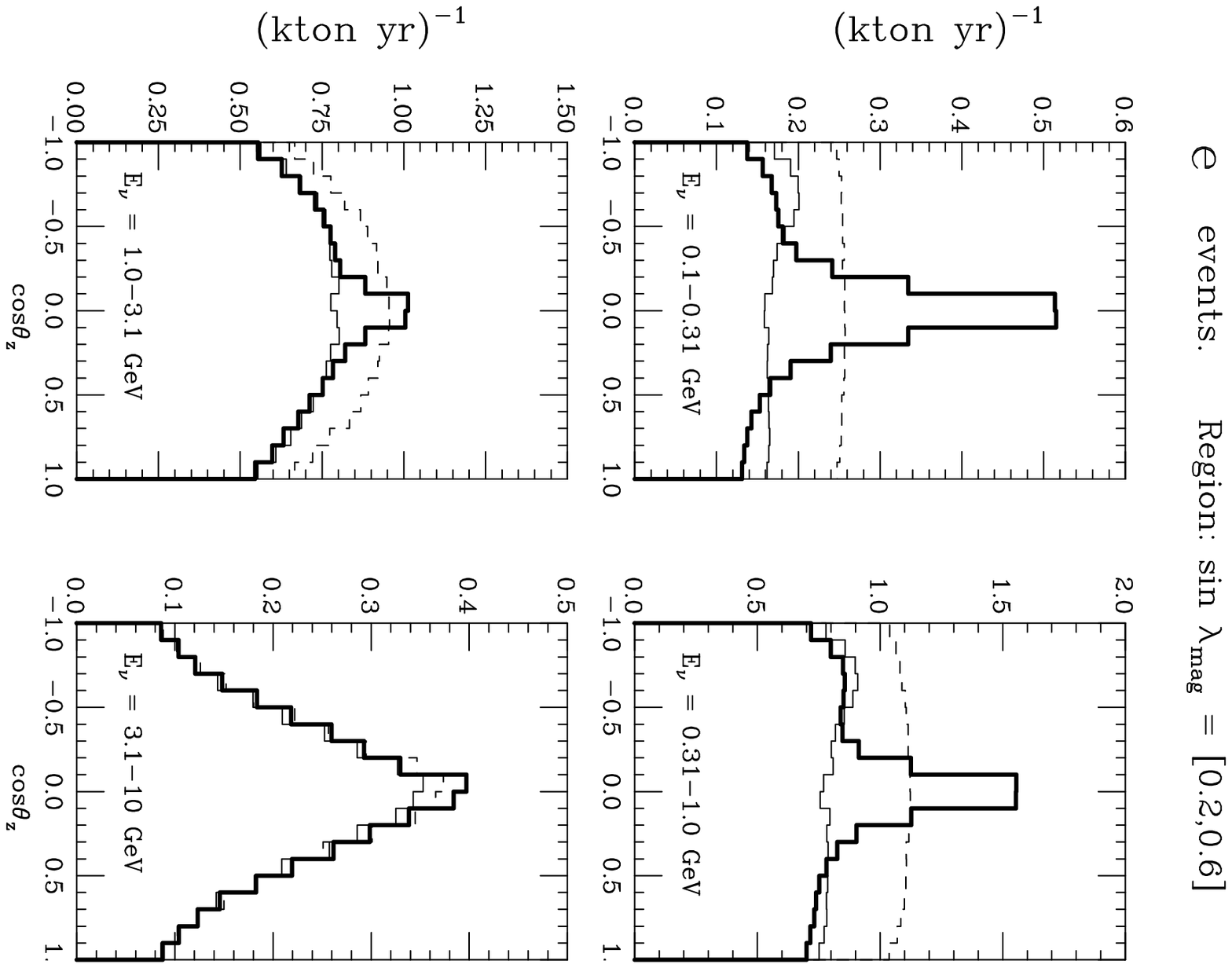,angle=90,height=18.0cm}}
\caption {Average zenith angle  distributions  for $e$--like events 
for  detectors located  in positions on the Earth with magnetic  latitude
$ \sin \lambda_M = [0.2,0.6]$.
The four panel  are for   different  neutrino energy intervals.
The  three histograms  are  for: fully 3--D calculation
(thick), 1--D calculation (thin),  1--D without geomagnetic  effects
(dashed).  
\label{fig:e_2}  }
\end{figure}

The   six figures  (from~\ref{fig:mu_1}  to~\ref{fig:e_3})
 with the results  of our calculation
are  organized as follows:
three  figures 
are for $\mu$--like  events  and three for
$e$--like  events;
in each  set  the  three  figures 
correspond  to the three different  magnetic  latitude  regions:
polar ($\sin \lambda_M = [0.6,1]$  where $\lambda_M$ is the magnetic
latitude),
intermediate ($\sin \lambda_M = [0.2,0.6]$)
and equatorial  ($\sin \lambda_M = [-0.2,+0.2]$).
 In each  figure   four panels  show  the  zenith
angle distributions for the  event  rates in   four  different
neutrino energy intervals.  To compute the rates  the
$\nu$ fluxes have been  convoluted  with the  neutrino cross sections model
described in  \cite{LLS}.  Note that  the zenith angle is the 
neutrino  one,  and  no  experimental  smearing or inefficiency
has been  included.

The motivation to perform the three  different  calculations are
that  a comparison between the three allows  to  put in  evidence
the  different     effects  that   determine the shape of  the zenith angle
distributions of the neutrino fluxes.
In fact:
\begin {enumerate}

\item In the calculation (iii) 
(1--D approximation with no geomagnetic  effects)
the   only   source of a non--flatness   in the
zenith angle dependence of  the 
event rates   are the neutrino yields. 

\item In calculation (ii) (1--D approximation) the zenith angle 
distributions  reflect both 
the geomagnetic effects on the primary cosmic rays, and the zenith angle
dependence of  the neutrino  yields.

\item  Finally  for    the full  three--dimensional  calculation (i) 
  the  zenith angle  distributions  
are  determined by a combination  of  all three effects:
the geomagnetic  effects, the neutrino
 yields and the spherical  geometry of
the source.
\end{enumerate}

Note that in  the calculation  (iii) where
geomagnetic  fields  are  neglected
and the primary cosmic  ray fluxes are considered  as 
isotropic the   neutrino event rates 
are independent from  the detector position,
therefore  the results   (described  by the dashed  histograms)
are equal (within the statistical errors of the montecarlo calculations)
for the same  event type ($\mu$ or $e$)  and  same energy range
in each of the   three regions 
that  correspond to  different  magnetic   latitude.
For the other two calculations   the   neutrino  event rates
do  depend  on the  detector  position and therefore the results are different
in the  three magnetic  latitude regions.

Let us consider   first the   zenith angle  distributions 
of  the  $\nu$ fluxes in  calculation (iii)
(1--D with no geomagnetic effects,  represented by the dashed  histograms 
in fig.~\ref{fig:mu_1}--\ref{fig:e_3}). 
For  $E_\nu \aprle 1$~GeV, the   $\nu$ fluxes
are  essentially isotropic, while  with   increasing energy
the fluxes  start to  develop an enhancement  for horizontal  directions.
This  enhancement  is due  to  the larger neutrino  yields in more
inclined  showers, and  becomes  more  marked  when  the 
energy increases as discussed in  section~\ref{sec:yield}.
Note that  the   enhancement   caused by the  neutrino
yields  is  more marked  for  $e$--like  events,
reflecting the  fact that  the effect of the shower
inclination is  especially important for muon  decay, that is the source
of essentially all  $e$--like events and  of  only a fraction 
(less or  equal  to one half)  of the $\mu$ events.
Note  also that this  effect is exactly  up--down symmetric,
and   all dashed  histograms  remain  equal (within  statistical errors)
for a reflection ($\cos \theta_z  \to -\cos \theta_z$).

The zenith angle  distributions  of calculation (ii)  
(1-D,  represented  by thin  histograms)
 very clearly display  the effect of  the geomagnetic
cutoffs.  This  is  reflected in  two  effects: the calculated event 
rates now  depend   on the detector  position and  are highest in
the polar region,  and  lowest   in the equatorial region
(the $y$ axis  is  an  absolute scale in fig.~\ref{fig:mu_1}--\ref{fig:e_3}),
 and the up--down symmetry  is broken.
For example  in the  calculations for the polar region
(fig.~\ref{fig:mu_1} and fig.~\ref{fig:e_1})  the up--going
($\cos  \theta_z < 0$) rates  are  smaller  that  the 
down--going ones ($\cos  \theta_z > 0$)   reflecting  the fact that
the   geomagnetic  cutoffs above  the detector  (in the  magnetic
polar region)
are lower  than  average, while  up--going  trajectories are
produced   everywhere on the Earth.
On the contrary    in the calculations
for  the equatorial region
(fig.~\ref{fig:mu_3} and fig.~\ref{fig:e_2})  the up--going
rates   are   larger  than the down--going ones    that are  suppressed
because of the  high  geomagnetic cutoffs in  the equatorial region.
The    geomagnetic    effects  become  negliglible  when the
neutrino  energy becomes  larger  than  a  few GeV,  reflecting the fact 
that higher energy  neutrinos  are  produced  in the showers  of 
higher energy  primary particles  that  are not  affected  by the geomagnetic
effects.

Finally  the  3--D calculation   (thick  histograms)
clearly  exhibits the  contribution from all three  sources 
of  zenith angle  dependence.
A new  and remarkable  feature is  present for  $E_\nu$  below few GeV's, and
is  the  sharp    enhancement for the  horizontal  directions,  whose origin
we  have discussed in section~5.  

It should be noticed  that the ``geometric''
enhancement for  the horizontal plane
in  a realistic calculation that  includes  the geomagnetic effects 
is {\em stronger}   than  what can be estimated
assuming that  the primary spectrum is  isotropic.
This  can be  easily understood  qualitatively  noting that the
value of the geomagnetic  rigidity cutoff (see eq.~\ref{eq:Stormer})
for  a  fixed position  and  azimuth   grows   monotonically 
with  zenith angle  from  a minimum value  for  the vertical direction
to a maximum value on the horizontal plane.
The enhancement on the horizontal plane,
 as discussed  in the previous  section,
can be understood as  the consequence of the fact that
there are more  ``vertical''  showers
producing  ``horizontal''  $\nu$'s
than vice versa.   This is true  for an isotropic  flux,
when the   source of the  vertical/horizontal  asymmetry is 
due to the fact  that the zenith angle distribution of
the showers is   $\propto \cos \theta_0$  (with $\theta_0$);
the asymmetry between vertical and horizontal  showers
becomes  more important when the geomagnetic effects are included,
since in this case the horizontal  primary 
flux  is also suppressed  by an higher rigidity cutoff.
For a realistic calculation the size of the ``geometric'' horizontal
enhancement   also  depends  on the position of the detector,
reflecting the effect  discussed above, and the  fact that the 
energy spectrum of the $\nu$ flux  observed in different location
changes being softer (harder) at high (low) magnetic  latitude,
and the angle $\langle \theta_{0\nu}\rangle$ is   correlated
with the neutrino energy.

The ``spherical  geometry'' enhancement
can be  easily distinguished  from the ``neutrino yield''  enhancement
since it  is  most important at low  neutrino energy reflecting
the   higher  average  neutrino--primary angle, and vanishes as
the angle $\theta_{0\nu}$   shrinks  with increasing $E_\nu$.  
Conversely the  ``neutrino yield''  enhancement  becomes more
important  as  the neutrino energy increases.

\section {Conclusions}
In this work  we  have discussed the  different  sources
that  together with neutrino oscillations  determine
the zenith angle distributions  of the atmospheric  neutrino fluxes.
The source volume of  the atmospheric  neutrinos  has the shape
of a spherical shell, and  a  simple  consequence of this
geometry is  the  existence of  
an  enhancement  of the flux intensity 
  for the  horizontal  directions.
This   enhancement  is   stronger  when the neutrinos are
emitted with  a large average angle  with  respect to the
primary particle  direction, and  therefore 
it is   most  important at low  energy, and  
disappears  for  $E_\nu \aprge 3$--10~GeV  when 
the neutrino emission is  approximtely collinear
with the shower axis.

The  detection of  this  enhancement is  not easy,  since the angular
resolution   for sub--GeV neutrinos  is  poor. 
This is   true for a  detector  such as Super--Kamiokande that can measure
only  the   charged lepton   produced in a  quasi--elastic  scattering
(such as $\nu_\mu n \to \mu^- p$), since the  angle
between the  neutrino and  the final  state  charged  lepton
is large,  but it is also true  \cite{fluka-3d}
for    higher resolution   detectors  (such as Soudan--2 or
Icarus)  that are capable  of  detecting the recoil protons.
 Even  in  this case  the angular resolution will be limited  by the fact
that  the   spectator  nucleons  (that  are undetected)
can  absorb  a non  negligible 3--momentum, and therefore
the vector  sum   of the  recoil nucleon and charged  lepton
momenta  does not    correspond exactly to the initial neutrino
momentum.

This  difficulty in  the  detection  of the enhancement
does not  imply that the effect  is  negligible 
 when   the atmospheric
neutrino  data  is  analysed   to extract information  about the
oscillation  parameters.
Several  effects   can be  significant.
The most important    one is 
 a  significant modification
of the pathlength distributions of the neutrino  events.
These  distributions are  controlled
by the $\nu$ zenith angle  distributions 
that can   very significantly distorted
by the geometrical effects discussed here.
Note also that   the  height distribution  of the  horizontal neutrinos
for a {\em fixed}  value of the zenith angle, is also 
modified  with respect to  a 
1--D  calculation.  For  example  
in a 3--D calculation most of the sub--GeV 
 horizontal  neutrinos  are actually  produced
in  less inclined (and  deeper developing showers)  after emission
at  large angles  with respect to the shower axis, and are therefore
produced  at lower altitude   with respect  to
the 1--D case.
This  effect is second order with respect to the distortion
of the zenith angle  distribution but must also  be taken into 
account.
As an example,  
even if the  detector  resolution is 
so poor that the shape  of the  neutrino  angular distribution is 
not measurable, 
in the presence of $\nu_\mu \leftrightarrow \nu_\tau$ oscillations
 the   pathlength  distribution determines  the average
(zenith  angle integrated)  suppression of the 
$\mu$--like event rate  for a given set of oscillation
parameters.

The geometrical  effects  can  also  be a   non negligible 
correction for the  calculation of the $\mu/e$ ratio.
The  enhancement  for the horizontal directions  depend in fact 
on the  average  angle  $\langle \theta_{0\nu} \rangle$ between
neutrino and primary particle  and on the
average altitude  of the neutrino production.
Both these quantities are  slightly  different
for neutrinos  produced directly in meson decay
or in the chain  decay of a muon, and are therefore 
also slightly different for electron and muon (anti)--neutrinos.

Also  the absolute observable rates  for  sub--GeV 
neutrino events  depend  on these effects.  The   absolute rate 
is not  important in the  determination of the oscillation 
parameters, but  the consistency of  data and prediction
also for the absolute rate is  certainly desirable.

The   improved calculation of the atmospheric  $\nu$
fluxes is now an active field  of  research,
strongly stimulated  by the great importance of   the discovery of
the existence of flavor transitions  for these
neutrinos.
New improved  calculations  will   be important
in the detailed  interpretation of the existing  and future 
data on atmospheric  neutrinos, and in the 
unbiased determinations of  the  $\nu$ 
oscillation parameters  (or other parameters describing the
new physics  of the flavor transitions).
New  measurements of the primary  cosmic  rays
have been recently  performed \cite{Caprice,Bess}, and  have to
be included  in a new fit  for the primary cosmic rays, and
at the same  a significant effort has  been dedicated to the
 developoment of  improved
modeling  of hadronic  interactions  in
cosmic  ray  showers.
Because of  these developments in the field
we postpone to a  future work a  quantitative discussion  
of  a new  estimate of the oscillation parameters.
The  newly  recognizied  geometric  effects  that we have  discussed
in this work   will be important in these future
calculations.

\vspace {2 cm}
\noindent {\bf Acknowledgments}
Special thanks to   Giuseppe  Battistoni  for  several very useful
discussions  and to  Takaaki Kajita    for  encouragement.
 I'm  grateful to Tom Gaisser, 
Teresa Montaruli and Todor Stanev for  reading  the manuscript
and  giving suggestions  for improvements.
I also  acknowledge useful  discussions  with Ed Kearns, Todd Haines
and M.~Honda.

\newpage

\end{document}